# Evidence, Calibration, and Stability: A Triadic Framework for Hypothesis Testing Under Model Uncertainty

Subir Hait

Department of Counseling, Educational Psychology, and Special Education

Michigan State University

ORCID: 0009-0004-9871-9677

August 2026

Abstract

Statistical tests are often asked to do too much. A single reported result is expected to describe what the observed data say, reassure readers about repeated-sampling behavior, and remain convincing when the working model is perturbed. Those tasks are connected, but they are not equivalent. Fisherian inductive inference and Neyman-Pearson decision theory clarify the first two; robust testing, sensitivity analysis, fragility measures, multiverse analysis, and distributional-stability methods speak to the third. I propose Evidence-Calibration-Stability (ECS) as a framework for keeping these roles separate while reporting them together. Evidence is post-data. Calibration belongs to the design or procedure. Stability is the post-data distance from the benchmark analysis to a conclusion-reversing perturbation within a declared model neighborhood. Full ECS support is conjunctive: a strong coordinate cannot rescue a failed one. For finite-dimensional affine perturbations, I derive an exact ellipsoidal stability radius. For smooth nonlinear margins, a uniform quadratic-remainder condition yields a certified lower bound over a declared neighborhood, showing when the affine formula is only a surrogate. I also establish coordinate invariance and a matrix extension for multiple claims, and distinguish confirmatory calibration from descriptive calibration profiles when prespecification is unavailable. Simulations for the one-sample t test and Student's historical sleep data show that the three coordinates can lead to different interpretations. ECS is a formal synthesis, not a claim that evidence, power, or robustness is itself new.

## 1. Introduction

Hypothesis testing has never had a single uncontested interpretation. Fisher (1925) used significance tests as tools of inductive inference: the realized data were judged for incompatibility with a null hypothesis. Neyman and Pearson (1933) built decision rules and evaluated them by long-run error frequencies and power. Lehmann (1993) argued that the traditions differ philosophically even when practice mixes them. That distinction still matters. A p value, a significance level, and a scientific conclusion are not interchangeable objects, even though applied reports often treat them as if they were.

The model itself creates another problem. Every test rests on assumptions about distribution, dependence, variance, measurement, sampling, or analysis. Misspecification can change both interpretation and operating behavior (White, 1982), and Box's (1976) warning about simplified models remains relevant. A small p value cannot certify the assumptions that produced it (Greenland et al., 2016; Wasserstein & Lazar, 2016; Wasserstein et al., 2019). The practical question is not just whether a result crosses a boundary. It is also whether the procedure deserved confidence and how far the conclusion can be pushed before it breaks.

The literature already contains serious answers to pieces of that question. Robust testing studies worst-case performance in neighborhoods of idealized distributions (Huber, 1965; Gao et al., 2018). Error-statistical work links post-data inference to severity and to statistical misspecification (Mayo & Spanos, 2006). In observational studies, sensitivity and robustness values quantify the hidden bias needed to change a conclusion (Zhao, 2019; Cinelli & Hazlett, 2020). The fragility index counts outcome changes that would reverse significance in randomized trials (Walsh et al., 2014). Multiverse and specification-curve analyses expose variation across defensible analytic choices (Steegen et al., 2016; Simonsohn et al., 2020). Distribution-shift work takes another route: the s-value measures parameter instability under Kullback-Leibler shifts (Gupta & Rothenhäusler, 2023), while distributionally robust inference seeks procedures that generalize across shifts (Rothenhäusler & Bühlmann, 2023).

These contributions make a broad claim of "adding robustness to hypothesis testing" untenable. The open organizational problem is narrower. A reader of a test result needs to distinguish at least three questions:

1. **Evidence:** What does the realized data analysis say relative to a declared evidential boundary?
2. **Calibration:** Did the procedure, as designed, meet acceptable error-control and power requirements for the scientific problem?
3. **Stability:** How far is the realized conclusion from reversal under a declared class of model or analysis perturbations?

ECS treats those questions as separate coordinates of one testing conclusion. I do not propose another omnibus score; collapsing the coordinates would erase the distinction the framework is meant to preserve. The evidential measure can vary, and so can the model neighborhood used for stability. What ECS supplies is an architecture for declaring those choices and keeping the resulting claims apart.

The contribution is deliberately narrower than a new test. ECS separates the temporal roles of evidence, calibration, and stability and draws a hard line between confirmatory calibration and post hoc operating-characteristic profiles. Full support is conjunctive, so strength on one coordinate cannot repair failure on another. On the technical side, the paper derives the exact affine ellipsoidal radius, a nonlinear remainder certificate, coordinate invariance, and a multiple-claim extension. The simulation and historical example show why these distinctions can change interpretation in practice.

ECS can be reported continuously or used confirmatorily when thresholds were fixed before analysis. I reserve the confirmatory version for settings in which a genuine binary or regulatory decision is needed. A result at 0.049 should not be treated as scientifically discontinuous from one at 0.051. The profile matters more than the label.

## 2. Three Distinct Statistical Tasks

### 2.1 Evidence from the realized data

Let $X$ denote observed data and let $\mathcal{M}_0$ denote the benchmark analysis model. Let $T(X;\mathcal{M}_0)$ be a statistic whose larger values favor a scientific claim, with a declared boundary $c$. A simple frequentist evidential margin is

$$E(X) = T(X;\mathcal{M}_0) - c.$$

The sign of E tells which side of the declared boundary the realized analysis occupies; its magnitude gives distance from that boundary on the chosen evidence scale. A likelihood ratio, e-value, Bayes factor, or another monotone measure could replace the test-statistic margin, and likelihood-based approaches offer an established alternative interpretation of evidence (Goodman, 1993; Royall, 1997). ECS takes no position on the single best evidence measure. The choice and its interpretation have to be declared.

The important point is temporal: $E(X)$ is a **post-data** quantity. It describes the realized sample. It is not the probability that the null hypothesis is true, and it is not the power of the design. Greenland et al. (2016) document how often those distinctions are lost in practice.

### 2.2 Calibration of the testing procedure

Calibration concerns the procedure before any particular dataset is observed. Let $\phi(X)$ denote the testing rule and let $\mathcal{G}_0$ be the prespecified class of data-generating distributions regarded as relevant under the null. Let $\mathcal{G}_1(\Delta^\star)$ represent scientifically important alternatives, where $\Delta^\star$ is a minimum effect or discrepancy worth reliably detecting. Define

$$\alpha_\varphi^{\max} = \sup_{P\in\mathcal{G}_0} P\{\varphi(X)=1\},$$

and

$$\pi_\varphi^{\min}(\Delta^\star) = \inf_{P\in\mathcal{G}_1(\Delta^\star)} P\{\varphi(X)=1\}.$$

Given a maximum tolerated Type I error $\alpha^\star$ and minimum required power $\pi^\star$, define the calibration margin

$$C(\phi) = \min\{\alpha^\star - \alpha_\phi^{\max},\ \pi_\phi^{\min}(\Delta^\star) - \pi^\star\}.$$

The testing procedure satisfies the declared calibration requirement when $C \geq 0$. The two components should ordinarily be reported alongside $C$ because they have different scientific meanings.

Observed power is not calibration. Plugging the realized effect estimate back into a power formula largely recycles the observed test statistic, so it does not provide an independent design check. Calibration should come from the design, exact operating properties, or a simulation plan defined without reference to the realized result.

### 2.2.1 Confirmatory and descriptive calibration

I take a strict view of confirmatory calibration: the Type I error limit, minimum acceptable power, and meaningful alternative must be fixed independently of the realized effect. Many observational, secondary-data, and historical analyses do not have that design history. In those cases the missing prespecification is simply missing. It should not be reconstructed from the observed effect or turned into a retrospective pass/fail label.

When prespecification is absent, the useful object is a descriptive calibration profile. Hold the design and analysis rule fixed, then report the operating characteristics over an externally justified range of alternatives. A simple version pairs the null rejection probability with the power curve as effect size varies.

$$K(\delta) = (\alpha\varphi, \pi\varphi(\delta)), \quad \delta \in \Delta.$$

That distinction changes the report. A secondary-data analysis can show $K(\delta)$, or a power table, without pretending that the observed effect was the original design target. The evidence and stability coordinates remain reportable, while the confirmatory scalar C is marked 'not prespecified.' No global conjunctive label is assigned. If several minimum effects have substantive justification, show them as scenarios rather than naming one retroactively as the design target.

## 2.3 Stability of the realized conclusion

Stability is again post-data, but it asks a different question from evidence. Let $\mathcal{U} \subseteq \mathbb{R}^q$ be a declared perturbation set, with $0 \in \mathcal{U}$ corresponding to the benchmark analysis. Let

$$m(X; u)$$

be a decision margin, defined so that $m(X; u) > 0$ supports the claim and $m(X; u) \leq 0$ does not. Let W be a symmetric positive-definite matrix defining the perturbation norm

$$\| u \|_W = (u^\top W u)^{1/2}.$$

For a benchmark analysis that supports the claim, $m(X; 0) > 0$, define the stability radius over the declared perturbation class by

$$S(X) = \inf\{ \|u\|_W : u \in \mathcal{U}, m(X; u) \leq 0 \}.$$

If m(X; 0) ≤ 0, set S(X) = 0 for that positively directed claim. If no perturbation in U reaches the failure set, use the standard convention inf ∅ = +∞. Thus S is the smallest normalized perturbation, within the declared class, that reverses the realized conclusion. The coordinates, the admissible set $\mathcal{U}$, and the metric W are part of the scientific specification; they are not tuning parameters hidden after the analysis.

The definition allows the perturbation coordinates to represent substantively different forms of uncertainty. Examples include an additive measurement offset, a standard-error inflation factor, a missingness deviation, a dependence parameter, or a finite set of alternative specifications. W can be Euclidean after meaningful scaling or can encode unequal and correlated uncertainty. The admissible set $\mathcal{U}$ may also impose bounds, sign restrictions, or other scientific constraints.

S is not a probability that the result is correct. It says nothing about how likely an adverse perturbation is. It answers a narrower question: how large must the declared departure become before the conclusion fails? That number is useful only when its units are scientifically interpretable, a point also raised in critiques of the fragility index.

## 3. The ECS Profile and the Triadic Adequacy Criterion

Define the ECS profile

$$\Psi(X, \phi) = \begin{pmatrix} E(X) \\ C(\phi) \\ S(X) \end{pmatrix}.$$

The vector notation is useful because it emphasizes that the three coordinates are not intended to be averaged into a single score. Let $e^\star$ be the evidential requirement, let calibration require $C \geq 0$, and let $s^\star$ be a minimum acceptable stability radius. The full-support region is

$$\mathcal{A}_{ECS} = \{(E, C, S): E > e^\star,\ C \geq 0,\ S > s^\star\}.$$

For a standard one-sided test, one may take $e^\star = 0$. The stability requirement $s^\star$ has no universal value; it must correspond to a scientifically meaningful perturbation scale. If the uncertainty coordinates are normalized so that $\| u \|_W = 1$ represents one jointly plausible unit of model departure, then $s^\star = 1$ has a direct interpretation: the conclusion survives every perturbation inside that unit neighborhood.

The conjunctive rule is important. Consider three hypothetical results:

| ECS profile | Interpretation |
| --- | --- |
| $E > 0, C \geq 0, S > s^\star$ | Evidentially positive, adequately calibrated, and stable |
| $E > 0, C \geq 0, S \leq s^\star$ | Positive and calibrated, but structurally fragile |
| $E > 0, C < 0, S > s^\star$ | Positive and stable, but produced by a design that failed the prespecified calibration requirement |

A large value of $E$ cannot make $C$ nonnegative. Similarly, an overpowered design does not make a fragile realized result stable. This noncompensatory structure is the reason ECS is a profile rather than an index.

For confirmatory settings, define component nulls

$$H_{0E}: E \le e^{\star}, \qquad H_{0C}: C < 0, \qquad H_{0S}: S \le s^{\star}.$$

The global null and alternative are

$$H_{0,ECS} = H_{0E} \cup H_{0C} \cup H_{0S},$$

and

$$H_{1,ECS} = H_{1E} \cap H_{1C} \cap H_{1S}.$$

The intersection-union logic makes the confirmatory label demanding by design: every required component must pass. That label is therefore reserved for analyses whose thresholds, including the calibration target, were fixed independently of the realized result. Exploratory and secondary-data work does not need a forced global verdict. Evidence and stability can be reported continuously, with calibration shown as an operating-characteristic profile instead of a retrospective pass/fail decision.

## 4. Exact Affine Stability Geometry and Nonlinear Margins

The abstract definition of S becomes especially transparent when perturbations affect the decision margin affinely. For the closed-form result below, suppose the admissible perturbation space is unconstrained, $\mathcal{U} = \mathbb{R}^q$, and

$$m(X; u) = m_0 - a^{\top} u,$$

where $m_0 = m(X; 0) > 0$ is the benchmark decision margin and $a \in \mathbb{R}^q$ collects the exact effects of the perturbation coordinates on the decision margin within the declared affine model class. No differential approximation is required for Proposition 1: the closed form is exact when the declared perturbation map is affine. If scientific restrictions make $\mathcal{U}$ a proper subset of $\mathbb{R}^q$, the general constrained optimization in Section 2.3 applies and the unconstrained affine optimizer must be checked for admissibility. This restriction is substantive. Contamination, dependence, missingness, measurement error, and many other scientifically relevant perturbations can produce nonlinear decision margins; in those settings the stability radius remains exactly defined by the optimization problem in Section 2.3, but the affine closed form need not equal it.

### Proposition 1. Ellipsoidal stability radius

Let W be symmetric positive definite and let $m(X; u) = m_0 - a^{\top} u$ with $m_0 > 0$ and $a \neq 0$. Then

$$S(X) = \frac{m_0}{\sqrt{a^{\top} W^{-1} a}}.$$

The conclusion-reversing perturbation of minimum W-norm is

$$u^\star = \frac{m_0 W^{-1} a}{a^\top W^{-1} a}.$$

Proof. Reversal occurs when $a^T u \geq m_0$. Because the W-norm is convex and the feasible set is a half-space, the minimum lies on the boundary $a^T u = m_0$. The generalized Cauchy-Schwarz inequality gives

$$(a^\top u)^2 \leq (a^\top W^{-1} a)(u^\top W u).$$

Therefore every reversing perturbation satisfies

$$\| u \|_W \geq \frac{m_0}{\sqrt{a^\top W^{-1} a}}.$$

Equality is attained by u* above. □

The formula separates two ingredients. The numerator $m_0$ is the observed margin from the decision boundary. The denominator describes how efficiently the declared model uncertainty can consume that margin. Two studies may therefore have the same benchmark evidence while having different stability radii because their conclusions respond differently to the same scientifically scaled perturbations.

### 4.1 Nonlinear margins: a uniform remainder certificate for the affine surrogate

For a smooth nonlinear decision margin, the affine calculation can still be useful, but only as a local surrogate whose error is controlled. Let $g$ be the gradient of the margin at the benchmark model and define $G = (g^\top W^{-1} g)^{1/2}$, the dual-norm size of that gradient. Suppose there is a declared radius $R > 0$ such that, for every perturbation with $\|u\|_W \leq R$, the nonlinear remainder is uniformly bounded by a quadratic term with constant $K \geq 0$:

$$|m(u) - m_0 - g^\top u| \leq (K/2)\|u\|_W^2, \quad \text{for } \|u\|_W \leq R.$$

**Lemma 1. Quadratic-remainder lower bound**

Assume $m_0 > 0$, $g \neq 0$, and that the quadratic remainder bound above holds throughout the radius-$R$ neighborhood. Then, for every $r \in [0, R]$ and every perturbation satisfying $\|u\|_W \leq r$,

$$m(u) \geq m_0 - Gr - (K/2)r^2.$$

Consequently, no conclusion reversal is possible at any radius $r \leq R$ for which the right-hand side is positive. For $K > 0$, define

$$S_L = [\sqrt{(G^2 + 2Km_0)} - G]/K, \quad \text{so that } S \geq \min\{R, S_L\}.$$

Proof. For $\|u\|_W \leq r$, generalized Cauchy-Schwarz gives $g^\top u \geq -Gr$. Combining this inequality with the quadratic remainder bound yields $m(u) \geq m_0 - Gr - (K/2)r^2$. If that lower bound is positive, reversal is impossible anywhere in the radius-$r$ ball. Solving $m_0$ − $Gr - (K/2)r^2 = 0$ gives the positive root $S_L$. The certificate is therefore valid through the smaller of the radius on which the remainder bound is known to hold and $S_L$. When $K = 0$, the same argument gives $S \geq \min\{R, m_0/G\}$; if the affine model holds globally, the exact radius is $m_0/G$. □

When K = 0 and the affine representation is valid on the full perturbation space, the bound reduces exactly to the affine radius $m_0/G$. For genuinely nonlinear margins, the quadratic remainder instead determines how much of the nominal affine radius can be certified without solving the full nonlinear optimization. Directional information about the second-order term can sharpen this conservative certificate, but that refinement is not required for the ECS architecture.

Use Proposition 1 literally only when the perturbation map is affine and the unconstrained optimizer is admissible. Otherwise solve the constrained or nonlinear problem, or report the affine value explicitly as a surrogate with an approximation certificate. Calling the local formula "exact" outside that setting would overstate what the geometry proves.

**Proposition 2. Coordinate invariance**

Let $u = Lv$ for an invertible matrix $L$, and transform the metric and sensitivity coefficients as

$$W_v = L^\top W_u L, \qquad a_v = L^\top a_u.$$

Then the ellipsoidal stability radius is unchanged:

$$\frac{m_0}{\sqrt{a_v^\top W_v^{-1} a_v}} = \frac{m_0}{\sqrt{a_u^\top W_u^{-1} a_u}}.$$

**Proof.** Since $(L^\top W_u L)^{-1} = L^{-1} W_u^{-1} L^{-\top}$,

$$a_v^\top W_v^{-1} a_v = a_u^\top L (L^\top W_u L)^{-1} L^\top a_u = a_u^\top W_u^{-1} a_u.$$

□

This is not a cosmetic invariance result. A stability radius should not change merely because the same uncertainty is written in different coordinates. Rescaling the coordinates without transforming the metric is a different scientific specification, not an equivalent reparameterization.

**4.2 Multiple conclusions and a matrix formulation**

Suppose an analysis contains $K$ claims whose benchmark decision margins are collected in

$$m_0 = \begin{pmatrix} m_{01} \\ \vdots \\ m_{0K} \end{pmatrix},$$

and suppose the declared perturbation $u \in \mathbb{R}^q$ acts through a $K \times q$ matrix $A$:

$$m(u) = m_0 - Au.$$

Assume first that every benchmark margin is positive; if any $m_{0j} \leq 0$, the joint conclusion has already failed and $S_{joint} = 0$. Let $a_j^T$ denote row $j$ of $A$. For a row with $a_j \neq 0$, Proposition 1 gives the distance to that claim's failure hyperplane; a row with $a_j = 0$ cannot be reversed by the declared affine perturbation and is assigned radius $+\infty$. Hence the radius to the first failed claim is

$$S_{joint} = \min_{\{j:\, a_j \neq 0\}} \ m_{0j} \,/\, \sqrt{(a_j^T W^{-1} a_j)},$$

This extension is useful for multiple endpoints, a family of contrasts, or another prespecified collection of claims. A records how the declared perturbations move each margin; W records how those perturbations are scaled. The joint radius is simply the distance to the first reachable failure hyperplane.

## 5. Why the Three Coordinates Do Not Collapse into One Another

The coordinates can move together without being interchangeable. A stronger observed margin often increases stability, and larger samples often improve evidence and power at the same time. ECS does not assume statistical independence or orthogonality. The claim is narrower: no one coordinate generally determines the other two.

**Proposition 3. Noncollapse of ECS coordinates**

Across admissible testing problems, there is no universal one-to-one transformation that recovers evidence, calibration, or stability from either of the remaining coordinates.

**Construction.** Fix a procedure and a realized benchmark decision margin $m_0 > 0$. Two analyses can have identical $E$ and $C$ yet use the same perturbation units with different exact perturbation effects, $m_A(u) = m_0 - u$ and $m_B(u) = m_0 - 2u$. Their stability radii are $m_0$ and $m_0/2$, respectively. Thus $E$ and $C$ do not determine $S$.

Conversely, the same observed standardized statistic can arise in studies designed with different sample sizes or variance assumptions. The realized evidence may therefore be identical while design power at a prespecified $\Delta^\star$ differs, showing that $E$ does not determine $C$. Finally, two studies may be constructed with the same calibration requirement and the same normalized stability radius but different benchmark evidence by changing the ratio between decision margin and perturbation effect while holding their quotient fixed. Thus $C$ and $S$ do not determine $E$. ▫

Proposition 3 is a logical claim, not an empirical correlation claim. ECS coordinates may be strongly associated in real data. Even then, they answer different questions and cannot in general be replaced by one member of the triplet.

**Proposition 4. Size control of a confirmatory ECS rule**

Suppose each stochastic component null used in a confirmatory ECS analysis is tested at level $\alpha$, and full ECS support is declared only when all component nulls are rejected and all deterministic design requirements are satisfied. Then the size of the global intersection-union test is at most $\alpha$.

**Proof.** Under $H_{0,ECS}$, at least one required component null is true or one deterministic requirement fails. If a deterministic requirement fails, global support is impossible. Otherwise, the probability of rejecting every component null cannot exceed the probability of rejecting the true component null, which is at most $\alpha$. No independence assumption among the component statistics is required for this bound. ▫

This result does not solve every inferential issue surrounding an estimated stability radius. Confidence bounds for $S$ can be nonregular near zero or when the perturbation model has corners. The proposition only states the logic of the conjunctive rule once valid component procedures are available.

## 6. Relationship to Existing Approaches

Table 1 places ECS beside neighboring approaches and makes the main overlaps visible.

**Table 1. Relationship of ECS to neighboring literatures**

| Approach | Primary target | Pre- or post-data? | Typical uncertainty object | Relationship to ECS |
|---|---|---|---|---|
| Neyman-Pearson testing | Type I/II error and power | Pre-data/procedure | Repeated samples under hypotheses | Basis of calibration coordinate |
| Severe testing | Evidential warrant using error probabilities | Pre- and post-data | Error probabilities and discrepancies | Closest philosophical bridge between evidence and calibration |
| Robust hypothesis testing | Worst-case performance of a rule | Primarily procedure-level | Distribution neighborhoods | Can be used to define design-class calibration |
| Fragility index | Reversal of significance in binary trials | Post-data | Changed event statuses | Special discrete reversal measure; narrower than general ECS stability |
| Sensitivity/robustness values | Reversal under hidden bias | Post-data | Causal sensitivity parameters | Domain-specific stability measures that can populate the ECS stability coordinate |
| Multiverse/specification curve | Variation across defensible analyses | Post-data | Alternative specifications | Can define a discrete model |

| Approach | Primary target | Pre- or post-data? | Typical uncertainty object | Relationship to ECS |
|---|---|---|---|---|
| | | | | neighborhood for stability |
| *s*-value | Parameter instability under distribution shift | Post-data | KL divergence | Closely related distributional-stability concept; ECS adds evidence and calibration architecture |

Robust hypothesis testing is the closest source of confusion. A robust test chooses or modifies a rule to control worst-case performance over an uncertainty set; ECS asks how far a particular realized conclusion lies from reversal. I do not treat those as substitutes. A rule can have good minimax properties while one observation sits close to its decision boundary, and a realized result can be far from reversal even when the study was poorly powered for the minimum effect that mattered. Misspecification theory gives the same warning from another direction: operating behavior under an incorrect model can differ from nominal behavior (White, 1982). When the uncertainty set or decision margin is nonlinear, ECS returns to the optimization definition in Section 2.3 rather than forcing Proposition 1 onto the problem.

The s-value of Gupta and Rothenhäusler (2023) is especially close to ECS stability because it measures distributional instability and can identify shifts that reverse a conclusion. I do not claim priority over that idea. The difference is architectural: ECS places stability beside a pre-data calibration coordinate and a post-data evidence coordinate, then uses a noncompensatory rule when all components are required.

The fragility index makes a related point in a discrete setting: significance can depend on a small number of changed binary outcomes. That simplicity is useful. Its scope is narrower, however, and the index can track the p value and sample size closely. ECS generalizes the distance-to-reversal question rather than the fragility-index calculation itself.

Multiverse and specification-curve analyses target uncertainty from defensible analytic choices. In ECS terms, those choices form a finite perturbation set. If a meaningful metric exists, one can report the smallest declared analytic change that reverses the claim. If no metric is defensible, the honest report is the worst-case margin across the finite set, not an artificial numerical radius.

## 7. ECS Implementation for the One-Sample t Test

To make the framework operational, consider a one-sided one-sample test

$$H_0: \mu \le 0 \qquad \text{versus} \qquad H_1: \mu > 0.$$

Let

$$t = \frac{\bar{X}}{s/\sqrt{n}},$$

and let $c = t_{n-1,1-\alpha}$. The evidential coordinate is

$$E = t - c.$$

For calibration, suppose the design requirement is size at most $.05$ and power at least $.80$ for a standardized effect $d^\star = 0.30$. Under the normal model, size is exactly $.05$ and power is obtained from the noncentral *t* distribution. For $n = 40, 71$, and $100$, nominal powers are approximately $.587$, $.805$, and $.909$, respectively. Thus the first design fails the declared calibration requirement, whereas the latter two pass.

For stability, suppose analysts are willing to stress-test two explicitly scaled departures from the benchmark analysis:

- an additive location offset of $b_0 = 0.10$ outcome standard-deviation units per unit of $u_1$; and
- a standard-error inflation of $r_0 = 10\%$ per unit of $u_2$.

Define $u = (u_1, u_2)^\top$ and use the Euclidean norm $\| u \|_2$. Under the perturbed analysis, the claim remains positive when

$$\bar{X} - b_0 u_1 - c\frac{s}{\sqrt{n}}(1 + r_0 u_2) > 0.$$

Rearranging gives the exact affine decision margin

$$m(u) = m_0 - a^\top u,$$

where

$$m_0 = \bar{X} - c\frac{s}{\sqrt{n}}$$

is the lower one-sided confidence-bound margin and

$$a = \begin{pmatrix} b_0 \\ c(s/\sqrt{n})r_0 \end{pmatrix}.$$

Proposition 1 therefore yields

$$S = \frac{m_0}{\sqrt{b_0^2 + \{c(s/\sqrt{n})r_0\}^2}}$$

whenever $m_0 > 0$. If $S > 1$, the positive conclusion survives every joint location/standard-error perturbation inside the unit circle defined by the declared reference scales. Importantly, the numerical value has meaning only because $b_0$ and $r_0$ have been stated.

## 8. Simulation Study

### 8.1 Design

The simulation examines how often conventional rejection and ECS support differ in a simple setting. We generated 30,000 independent samples for each combination of sample size $n \in \{40,71,100\}$, standardized mean effect $\mu \in \{0,.15,.30,.50\}$, and three error distributions: standard normal, a standardized Student $t_3$, and a standardized lognormal distribution with log-scale standard deviation $.75$. All error distributions were centered at zero and scaled to variance one.

The benchmark analysis was the one-sided one-sample t test at $\alpha = .05$. Calibration used the benchmark normal-model design requirement of power at least .80 for $d\star = .30$. Stability used $b_0 = .10$, $r_0 = .10$, and threshold $s\star = 1$. Thus a simulated result received full ECS support only if (a) the benchmark test rejected, (b) the benchmark design met the calibration requirement, and (c) the stability radius exceeded one reference perturbation unit. The calibration rule was fixed from the benchmark normal design and was not recalculated under the nonnormal stress scenarios; those scenarios evaluate the behavior of the prespecified procedure under distributional departure rather than redefine the calibration target after seeing the data-generating condition.

The point is diagnostic, not procedural. The simulation is meant to expose different classifications, not to advertise an optimal test.

### 8.2 Results under the normal model

Table 2. Selected conventional rejection and ECS support under normal errors

| $n$ | $\mu$ | Nominal power at $d^\star = .30$ | Conventional rejection | Stable among rejections ($S > 1$) | Full ECS support |
|---|---|---|---|---|---|
| 40 | 0.00 | .587 | .050 | .229 | .000 |
| 40 | 0.30 | .587 | .590 | .577 | .000 |
| 40 | 0.50 | .587 | .924 | .857 | .000 |
| 71 | 0.00 | .805 | .049 | .134 | .007 |
| 71 | 0.30 | .805 | .810 | .630 | .510 |
| 71 | 0.50 | .805 | .994 | .957 | .951 |
| 100 | 0.00 | .909 | .051 | .085 | .004 |
| 100 | 0.30 | .909 | .906 | .691 | .626 |
| 100 | 0.50 | .909 | 1.000 | .990 | .990 |

The $n = 40$ design illustrates the noncompensatory logic most clearly. At $\mu = .50$, the benchmark test rejects in more than 92% of samples and most rejections are stable under the declared model neighborhood. Nevertheless, the design fails the prespecified $.80$ power requirement at $d^{\star} = .30$, so full ECS support is not available. This is not a statement that the observed results are false. It is a statement that the study did not meet the declared confirmatory design standard.

Once the design is calibrated, stability does most of the extra filtering near the scientific boundary. At n = 71 and μ = .30, conventional rejection occurs in about .810 of samples but full ECS support in .510. By μ = .50, the gap is small because nearly every positive result is also beyond the stability threshold. The μ = .15 condition is omitted from Table 2 for space and appears in Figure 1 with the three tabulated effects.

***Figure 1***

*Conventional rejection and full ECS support for the calibrated n = 71 normal-design scenario.*

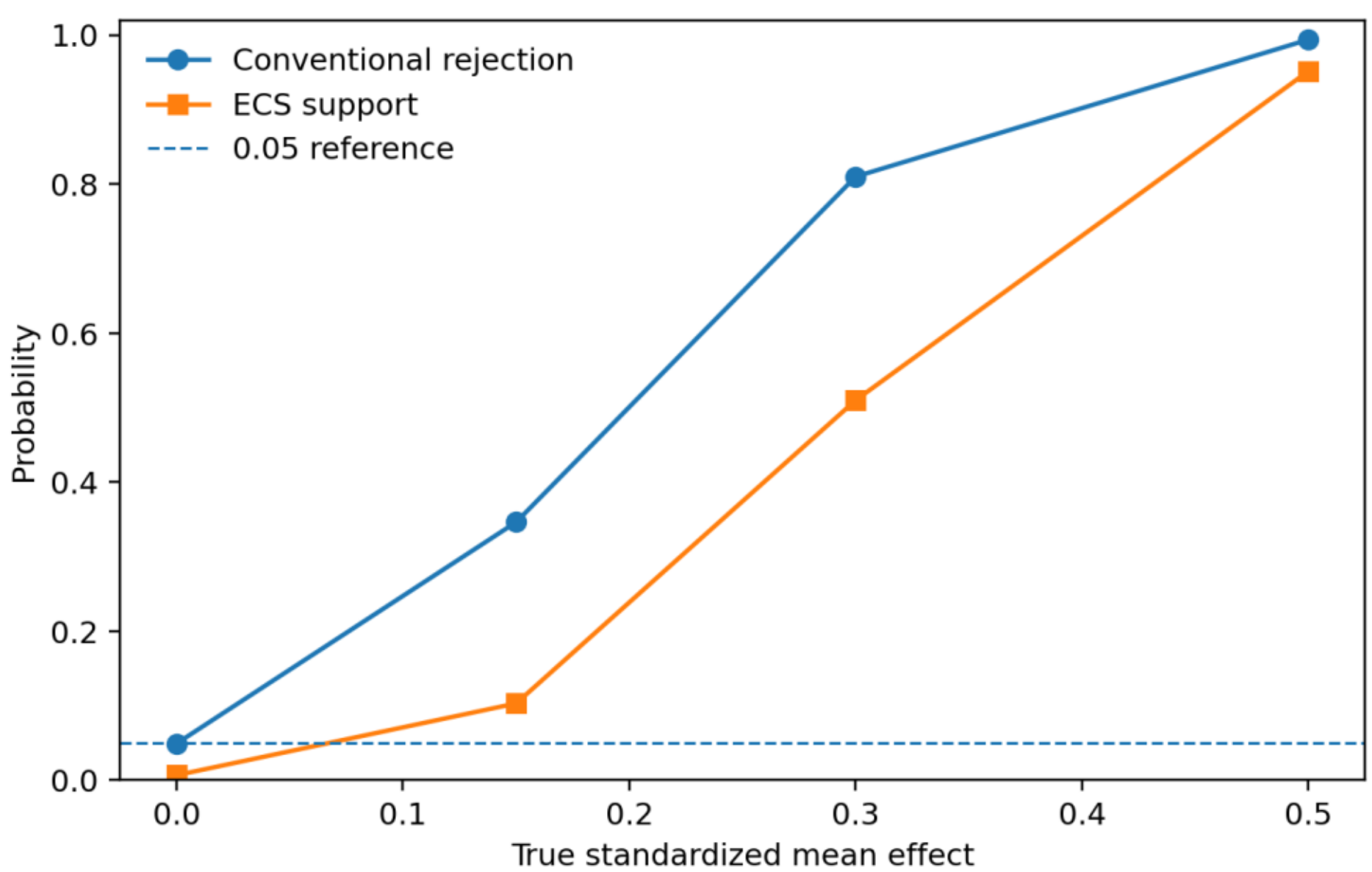

The null results also illustrate that the stability criterion acts as an additional filter rather than inflating false support. With $n = 71$, the conventional null rejection rate is .049, whereas full ECS support occurs in .007 of null samples under the declared stability threshold.

### 8.3 Distributional departures

At $n = 71$ and $\mu = .30$, the conventional rejection probabilities were .810, .835, and .891 under normal, $t_3$, and lognormal errors, respectively. The proportions of conventional rejections that also exceeded the stability threshold were .630, .682, and .568, producing full ECS-support probabilities of .510, .569, and .506. The framework therefore does not mechanically reward distributions that happen to increase nominal rejection. Stability remains a separate realized-data property.

Under the null, the one-sided $t$ test was close to nominal for normal and standardized $t_3$ errors (.049 and .052) and conservative for the chosen lognormal distribution (.021). ECS support under the null was .007, .004, and .002, respectively. These results should not be interpreted as universal robust-size guarantees; they simply illustrate the behavior of this particular implementation.

***Figure 2***

*Distribution of stability radii among conventional rejections for n = 71 and μ = .30. The dashed line marks the prespecified stability requirement S = 1.*

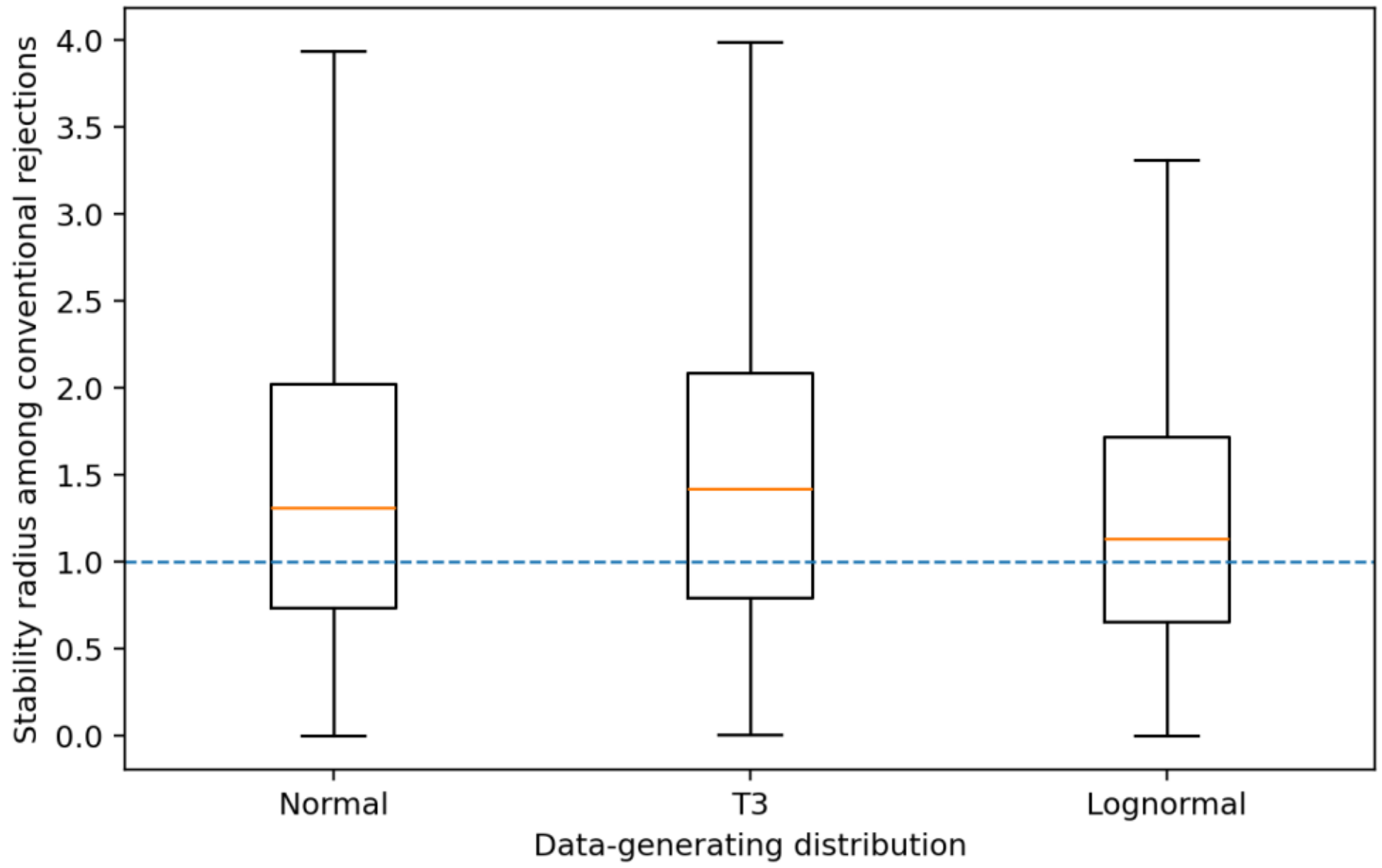

**8.4 Bootstrap uncertainty for the stability radius**

Because $S$ is estimated from the same data that determine the benchmark test, reporting a point estimate alone can overstate precision. As a preliminary check, we evaluated a nonparametric percentile bootstrap for the population analogue of the location/scale stability radius under normal sampling. Across four regular scenarios ($n = 71$ or 100 and $\mu = .30$ or $.50$), 95% percentile intervals achieved empirical coverage between .934 and .945 in 1,000 Monte Carlo repetitions with 300 bootstrap resamples per dataset. Coverage was therefore close to, but somewhat below, nominal.

The result is encouraging only as a baseline. Near $S = 0$, or when a perturbation set has corners or discrete model choices, ordinary bootstrap procedures may be nonregular. A mature implementation should report bootstrap diagnostics and use one-sided lower confidence bounds when stability is part of a confirmatory criterion.

**9. Historical Illustration: Student's Sleep Data**

The sleep dataset distributed with R records increases in hours of sleep for 10 patients under two soporific drugs. Cushny and Peebles (1905) reported the data, and Student (1908) used them in the development of the t test. For the paired differences, drug 2 minus drug 1, the mean is 1.58 hours and the standard deviation is 1.23 hours.

For the one-sided paired $t$ test of a positive mean difference,

$$t = 4.062,$$

with 9 degrees of freedom and one-sided $p = .0014$. The $.05$ critical value is 1.833, giving an evidential margin

$$E = 4.062 - 1.833 = 2.229.$$

For an illustrative stability analysis, suppose a modern analyst declares a reference additive offset of $.25$ hour and a 10% standard-error inflation. The one-sided lower-bound margin on the outcome scale is

$$m_0 = 1.58 - 1.833(0.389) = 0.867 \text{ hours.}$$

The exact ellipsoidal stability radius is then

$$S = \frac{0.867}{\sqrt{0.25^2 + \{1.833(0.389)(0.10)\}^2}} = 3.33.$$

Under these explicitly chosen scales, more than three reference perturbation units are required to reverse the positive conclusion. If the reference additive offset is doubled to $.50$ hour while the 10% standard-error scale is unchanged, the radius falls to approximately 1.72. The example makes the dependence of $S$ on a declared uncertainty geometry visible rather than implicit.

Calibration is where the historical example becomes deliberately incomplete. The original report supplies no modern prespecified minimum effect, target power, or design-class error requirement. That missing design history cannot be repaired after the fact. A power curve for externally chosen effects is still useful, but it is descriptive operating-characteristic information, not reconstructed prespecification. The ECS report should therefore mark C as 'not prespecified,' report evidence and stability, and add K(δ) only as a descriptive profile. No global conjunctive label is assigned.

$$\Psi = (E = 2.229,\ C = \text{not prespecified},\ S = 3.33 \text{ under the stated perturbation scale}).$$

The missing coordinate is informative. Strong evidence and apparent model stability do not manufacture design information. Secondary-data users can still show what the procedure would detect over scientifically interpretable alternatives, but the line between that description and genuine prespecification should remain visible.

**10. Practical Reporting**

A minimal ECS report should contain six items:

1. the scientific claim and benchmark statistical model;
2. the evidence measure and its continuous value;
3. for confirmatory analyses, the prespecified Type I error and minimum-power target, including the meaningful alternative; for secondary analyses, a descriptive calibration curve or table over externally justified alternatives;
4. the model or analysis perturbations used for stability;

5. the perturbation metric or reference scales;
6. the ECS profile, marking confirmatory calibration as “not prespecified” when appropriate, and using a global support label only when all thresholds were fixed independently of the realized result.

A compact report could read:

> The benchmark one-sided test gave $t = 2.41$ against a critical value of 1.67, so the evidential margin was $E = .74$. The design controlled size at .05 and had .86 power at the prespecified minimum effect, satisfying the calibration requirement. Under a model-neighborhood specification in which one perturbation unit corresponds to a .10-SD additive offset or a 10% standard-error inflation, the estimated stability radius was $S = 1.42$ (bootstrap lower bound 1.08). Because the prespecified stability requirement was one unit, the result satisfied all three ECS criteria.

That report says more than 'statistically significant' without pretending that ECS establishes scientific truth. It also tells the reader exactly what gives the stability number its meaning.

For a secondary-data analysis, a compact report might instead state: “The benchmark analysis produced evidential margin E = 0.84. No minimum effect and target power were prespecified for this dataset, so confirmatory calibration is not scored; the procedure’s power curve is reported over the externally chosen effect range 0.20–0.50. The estimated stability radius is S = 1.31 under the declared perturbation geometry. We therefore report an E–S result with descriptive calibration and do not assign a global ECS-support label.”

## 11. Limitations and Open Problems

Several limits are built into ECS.

Stability cannot be defined without a perturbation class and scale. A number such as $S = 2$ is meaningless unless the reader knows what two units represent. This is a feature as well as a limitation: the framework forces analysts to make model uncertainty explicit. In some applications, an ordered metric may be scientifically unjustified. A finite worst-case specification report is then preferable to an artificial continuous radius.

Evidence is still philosophically contested. The signed test-statistic margin is useful for exposition, but it does not settle disputes over p values, likelihood evidence, Bayesian evidence, or severity. ECS can accommodate different evidence coordinates. Comparisons across those paradigms still require care.

Calibration depends on the scientific alternative and on when that alternative was chosen. Power at an effect selected after seeing the data is not a meaningful design credential. For confirmatory use, the minimum effect must come from substantive considerations independent of the realized result. When that history does not exist, a descriptive calibration profile is the more honest report.

Evidence and stability can be strongly correlated. A result far from the benchmark boundary will often require a larger perturbation to reverse. I do not regard that dependence as a flaw. The relevant question is whether stability contains information not fixed by the evidential margin; Proposition 3 shows that it can.

Nonlinear stability is harder than the affine formula suggests. Proposition 1 is exact only for affine margins with an admissible optimizer. Smooth nonlinear problems can use the Section 4.1 certificate, but contamination, dependence, missingness, measurement error, and other constrained settings may need direct optimization. Inference is harder too. Corners, radii near zero, discrete specification sets, or nonunique nearest failure points can produce nonregular sampling distributions.

The conjunctive rule is intentionally conservative. A low-powered historical study can fail calibration even if an extreme result later replicates. I am comfortable with that consequence because ECS is judging whether a study met a declared three-part standard, not whether nature ultimately contains an effect. Exploratory work can avoid the global label and report the coordinates continuously.

## 12. Discussion

ECS starts from a simple objection: a test result is often asked to carry more meaning than any one statistic can support. What the realized sample says, how the procedure behaves across repetitions, and how dependent the conclusion is on the maintained model are different objects. Fisherian significance and Neyman-Pearson error control already separate the first two. Modern robustness and sensitivity work adds the model-dependence question. ECS keeps all of those distinctions visible in one report.

The practical recommendation is simple. Report evidence, calibration, and stability separately. Evidence describes the realized benchmark analysis. Calibration records the operating requirement fixed for the procedure, or remains descriptive when that requirement was not prespecified. Stability gives the distance to failure under the declared perturbation geometry. Do not average them.

The geometry gives the framework some teeth. Under an affine perturbation map, stability is an exact Mahalanobis-type distance to a failure hyperplane. Once the map becomes nonlinear or scientifically constrained, that convenience disappears: the correct object is the constrained distance to the failure set. The quadratic-remainder result certifies part of the local radius when its bound is valid, but direct nonlinear optimization is preferable when the perturbation map can actually be evaluated. Coordinate invariance and the multiple-claim result survive within the affine setting.

The simulations show where ECS changes interpretation. Conventional rejection can be common even when the design misses its prespecified power target. In calibrated designs, many near-boundary rejections still fall inside a one-unit uncertainty neighborhood. At larger effects the two decisions converge, as they should. ECS is most informative where the usual sampling boundary and the model-uncertainty boundary are both close.

I would not defend ECS as a new theory of evidence, power, or robustness. That claim would be too broad and the existing literatures are too mature. The defensible novelty is the formal architecture: it separates temporal roles, refuses compensation across failed requirements, and attaches an explicit stability geometry to ordinary testing procedures.

The next useful step is implementation. Software should accept a benchmark analysis, either a prespecified calibration target or a descriptive calibration grid, and a user-defined perturbation map, then return the ECS profile with diagnostics and uncertainty intervals. It also needs to say when a radius is exact and when it is only an approximation. The theory still has unfinished pieces. Sharper nonlinear certificates and principled model neighborhoods are two of them; uncertainty for multiple radii is another. Those are substantive problems, not packaging details.

The point is modest. Ask the sample question, the design question, and the model-dependence question separately. If calibration was not prespecified, say so. Evidence and stability can still be reported. That is better than inventing a retrospective pass/fail standard or asking one statistic to carry more meaning than it can support.